\documentclass[12pt]{article}                              %
\title{Einstein's Equations and Equivalent Hyperbolic      %
       Dynamical Systems                                   %
       \footnote{Based on a lecture given by James         %
                 W. York Jr. at the $2^\mathrm{nd}$        %
                 Samos meeting, 1 September, 1998.}}       %
\author{Arlen~Anderson\footnote{Department of Physics      %
        and Astronomy, University of North Carolina,       %
        Chapel Hill, NC 27599-3255 USA}, Yvonne            %
        Choquet-Bruhat\footnote{Gravitation et Cosmologie  %
        Relativiste, t.22-12, Un. Paris VI,                %
        Paris 75252 France},\\                             %
        and James W. York Jr.$^\dag$}                      %
\date{August 5, 1999}

\usepackage{latexsym}
\usepackage{amssymb}
\newcommand{\bpz}{\bar{\partial}_0}
\newcommand{\barbox}{\bar{\Box\mbox{\small\mathstrut}}}

\begin{document}

\maketitle

\begin{abstract}
We discuss several explicitly causal hyperbolic 
formulations of Einstein's dynamical $3+1$ equations
\index{Einstein equations!hyperbolic} 
in a coherent way, 
emphasizing throughout the fundamental role of the ``slicing
function\index{slicing function},'' 
$\alpha$---the quantity that relates the lapse\index{lapse}
$N$ to the determinant of the spatial metric 
\index{metric!spatial} $\bar{g}$ through
$N = \bar{g}^{1/2} \alpha$.  The slicing function allows us to 
demonstrate explicitly that every foliation of spacetime
by spatial time-slices can be used in conjunction with the 
causal hyperbolic forms of the dynamical Einstein equations.
Specifically, the slicing function plays an essential role
(1) in a clearer form of the canonical action 
principle\index{action principle!canonical} and
Hamiltonian dynamics\index{Hamiltonian dynamics} 
for gravity and leads to a recasting (2)
of the Bianchi identities\index{Bianchi identities}
$\nabla_\beta G^\beta\mathstrut_\alpha \equiv 0$
as a well-posed system for the evolution of the gravitational
constraints\index{constraint} in vacuum, and also (3) of 
$\nabla_\beta T^\beta\mathstrut_\alpha \equiv 0$ 
as a well-posed system for evolution of the energy and
momentum components of the stress tensor in the presence of
matter,  (4) in an explicit rendering of four hyperbolic 
formulations of Einstein's equations\index{Einstein equations!hyperbolic} 
with only physical
characteristics, and (5) in providing guidance to a new
``conformal thin sandwich'' 
form of the initial value constraints.
\index{constraint!initial value!conformal thin sandwich}
\end{abstract}

\section{Introduction}
Einstein's equations\index{Einstein equations} 
have, for much of the history of general
relativity, been explored very fruitfully in terms of their
concise and elegant statements characterizing the geometry
of four dimensional pseudo-riemannian geometries. Such
geometries depict possible physical spacetimes containing
only ``the gravitational field itself.'' The variety
and properties of these ``empty'' spacetimes is truly
astonishing. Quasi-local geometric entities such as trapped 
surfaces and event horizons have become familiar. 
It is now firmly established that large-scale
topological and geometrical features
of spacetime are, indeed, subjects
of physical inquiry. The nature and
distribution of ``matter'' at stellar scales and upward
has also brought particle physics and hydrodynamics to the
fore.

During these years, however, a steady development of the
``space-plus-time'' or $3+1$ view of spacetime geometry
has also occured. Here one views general relativity
as ``geometrodynamics''
\index{geometrodynamics} in the parlance of John
Wheeler~\cite{Wheeler}. The emphasis, in the canonical 
or Hamiltonian
explication of geometrodynamics\index{geometrodynamics}
\index{Hamiltonian dynamics} given by Arnowitt,
Deser, and Misner (``ADM'')~\cite{ADMDir} and by 
Dirac~\cite{Dir58,Dir59},
is on the evolving intrinsic and extrinsic geometry 
of spacelike hypersurfaces which determine, by knowledge
of the appropriate initial data and by classical 
causality, the spacetime ``ahead'' (and ``behind''),
if spacetime is globally hyperbolic, an assumption
we adopt throughout.

Underlying and preceding geometrodynamics
\index{geometrodynamics} and Hamiltonian
\index{Hamiltonian dynamics}
methods, however, was the basic realization that four 
of the ten Einstein vacuum equations\index{Einstein equations!vacuum}
are nonlinear
constraints\index{constraint!nonlinear} 
on the initial Cauchy data\index{Cauchy data}, 
which play
such a decisive role in defining the later canonical 
formalism~\cite{Lic}. The Cauchy problem\index{Cauchy problem}, 
constraints plus
evolution, was shown to be well-posed in the modern
sense of nonlinear partial differential 
equations~\cite{FB52,CB56,Lic44,CBY80}. 
This train of progress was
marked by early work of Darmois~and Lichnerowicz~\cite{Lic44},
and brought to
definitive development by one of us~\cite{FB52,CB56}.

At this writing, with accurate three-dimensional simulations
using the full Einstein equations, 
\index{Einstein equations} with and without the
presence of stress-energy sources, becoming essential for
realistic studies of gravity waves, high energy astrophysics,
and early cosmology, studies of Einstein's equations
of evolution in $3+1$ form have blossomed~[11-28]
(an incomplete sample --- see also~\cite{AY99}).
Hyperbolic forms, especially first-order symmetrizable forms 
possessing only physically causal directions of propagation,
have undergone very significant development in the past 
few years.

Specifically, in this chapter we describe in detail
several explicitly causal hyperbolic formulations of
Einstein's dynamical $3+1$ equations
\index{Einstein equations!hyperbolic} by following a
path that can be viewed as lighted by the
``slicing function,''
$\alpha$ --- the quantity that relates the lapse $N$\index{lapse}
to the determinant of the spatial metric $\bar{g}$
\index{metric!spatial}
through $N = \bar{g}^{1/2} \alpha$. This representation of the
lapse function was presented in~\cite{CBR83}.
The slicing function
allows us to demonstrate explicitly that no foliation of spacetime by
spatial time-slices can be an obstacle to the causal
hyperbolic forms of the dynamical Einstein equations. 
\index{Einstein equations!hyperbolic} The slicing
function plays an essential role (1) in a more precise form of 
the canonical action principle\index{action principle!canonical} 
and canonical dynamics for
gravity, (2) leads to a recasting of the Bianchi identities
\label{Bianchi identities}
$\nabla_\beta G^\beta\mathstrut_\alpha \equiv 0$ as a 
well-posed system for the evolution of the gravitational
constraints\index{constraint!vacuum} in vacuum and also (3) of $\nabla_\beta
T^\beta\mathstrut_\alpha \equiv 0$ as a well-posed system for
evolution of the energy and momentum components of the stress
tensor in the presence of matter, (4) in an explicit display
of four hyperbolic formulations of Einstein's equations
\index{Einstein equations!hyperbolic} with only
physical characteristics, and (5) even in providing guidance to a new
elliptic ``conformal thin sandwich'' form of the initial value
constraints\index{constraint!initial value!conformal thin sandwich}. 

We recall that the proof of the existence of a causal evolution in local
Sobolev spaces of $\bar{g}$ and its extrinsic curvature (second
fundamental tensor) $K$ into an Einsteinian spacetime does not result
directly from the equations giving the time derivatives of $\bar{g}$
and $K$ in terms of space derivatives of these quantities in a
straightforward $3+1$ decomposition of the Ricci tensor of the
spacetime metric, which contains also the lapse and shift
characterizing the time lines. These equations do not appear as a
hyperbolic system for arbitrary lapse and shift, in spite of the fact
that their characteristics are only the light cone and the time 
axis~\cite{Fri96}.

We now turn to notational matters, conventions, and to the $3+1$
decomposition of the Riemann and Ricci tensors. We assume here and
throughout the sequel that the spacetime $V=M\times\mathbb{R}$ is
endowed with a metric $g$ of signature $\left( -, +, +, + \right)$ and
that the time slices are spacelike, that is, have signature
$\left( +, +, + \right)$. These assumptions are not restrictive for
globally hyperbolic\index{globally hyperbolic} (pseudo-riemannian)
spacetimes. 

We choose on $V$ a moving coframe\index{moving coframe} such that the
dual vector frame has a time axis orthogonal to the slices $M_t$ while
the space axes are tangent to them. Specifically, we set
\begin{eqnarray}
\theta^0 &=& d t \; , \nonumber\\
\theta^i &=& d x^i + \beta^i dt \; ,
\label{basis}
\end{eqnarray}
with $t\in\mathbb{R}$ and $x^i$, $i=1,2,3$ local coordinates on
$M$. The Pfaff or convective derivatives $\partial_\alpha$ with
respect to $\theta^\alpha$ are
\begin{eqnarray}
\partial_0 &\equiv& \frac{\partial}{\partial t} - \beta^i \partial_i
 \nonumber\\
\partial_i &\equiv& \frac{\partial}{\partial x^i}
\label{co-basis}
\end{eqnarray}
In this coframe, the metric $g$ reads
\begin{equation}
ds^2 = g_{\alpha \beta} \theta^\alpha \theta^\beta
\equiv -N^2 (\theta^0)^2 + g_{i j} \theta^i \theta^j \; .
\end{equation}
The $t$-dependent scalar $N$ and space vector $\beta$ are
called the lapse function and shift vector of the slicing.
These quantities were explicitly identified in~\cite{CB56}
and play prominent roles in all subsequent $3+1$ formulations.
Any spacetime tensor decomposes into sets of time dependent
space tensors by projections on the tangent space or the normal to
$M_t$. 

We define for any $t$-dependent space tensor $T$ another such tensor
of the same type, $\bpz T$, by setting
\begin{equation}
\bpz \equiv \frac{\partial}{\partial t} - \pounds_\beta \; ,
\end{equation}
where $\pounds_\beta$ is the Lie derivative on $M_t$ with 
respect to $\beta$.

Notice that in our foliation-adapted basis\index{foliation-adapted 
basis}~(\ref{basis}) and~(\ref{co-basis}), that if $g$ 
denotes the spacetime metric
and $\bar{g}$ the space metric, then we have 
($g_{0 i} = g^{0 i} = 0$ in our frames):
\begin{equation}
g_{i j} = \bar{g}_{i j} \; ; \quad g^{i j} = \bar{g}^{i j} \; .
\end{equation}
(Greek indices range $\{0,1,2,3\}$ while latin ones are
purely spatial.) Hence, {\em no overbars will be used to denote
components of the spatial metric}. On the other hand, for
the determinants, we have $(-\det g) = N^2 (\det \bar{g})$.
Therefore we shall use overbars on spatial metric determinants;
for example $\bar{g}^{1/2} \equiv (\det \bar{g})^{1/2}$.

Likewise, to distinguish the purely spatial components of the
spacetime Ricci tensor (say), we shall write $R_{i j}(g)$,
while for the space Ricci tensor we shall write $R_{i j}(\bar{g})$.
In general, of course, $R_{i j}(g) \neq R_{i j}(\bar{g})$. The
Levi-Civita connection of $g$ is denoted by $\nabla$ and that
of $\bar{g}$ by $\bar\nabla$.

With the convention
\begin{equation}
\nabla_\alpha \sigma_\beta \equiv \partial_\alpha \sigma_\beta
     - \sigma_\rho \gamma^\rho\mathstrut_{\beta \alpha} \; ,
\end{equation}
and the definitions
\begin{eqnarray}
\gamma^\alpha\mathstrut_{\beta \gamma} &=& 
     \Gamma^\alpha\mathstrut_{\beta \gamma} + g^{\alpha \delta}
     C^\varepsilon\mathstrut_{\delta (\beta} g_{\gamma) \varepsilon}
     - \frac{1}{2} C^\alpha\mathstrut_{\beta \gamma} \; , \\
d\theta^\alpha &=& -\frac{1}{2} C^\alpha\mathstrut_{\beta \gamma}
     \theta^\beta \wedge \theta^\gamma \; ,
\end{eqnarray}
we have for the connection coefficients ($\Gamma$ denotes an 
ordinary Christoffel symbol)
\begin{equation}
\gamma^i\mathstrut_{j k} = \Gamma^i\mathstrut_{j k}(g)
     = \Gamma^i\mathstrut_{j k}(\bar{g}) \\
\end{equation}
\begin{equation}
\gamma^i\mathstrut_{0 k} = - N K^i\mathstrut_k \;, \quad
     \gamma^i\mathstrut_{j 0} = - N K^i\mathstrut_k 
     + \partial_j \beta^i \;, \quad
     \gamma^0\mathstrut_{i j} = - N^{-1} K_{i j}
\end{equation}
\begin{equation}
\gamma^i\mathstrut_{0 0} = N \partial^i N \; , \quad
     \gamma^0\mathstrut_{0 i} = \gamma^0\mathstrut_{i 0}
     = \partial_i \log N \; , \quad
     \gamma^0\mathstrut_{0 0} = \partial_0 \log N \; .
\end{equation}
Observe that if $\alpha$ is a space scalar of weight $-1$,
we have
\begin{eqnarray}
\bar\nabla_i \alpha &=& \partial_i \alpha + \alpha 
     \Gamma^k\mathstrut_{k i}(g) \,=\, \partial_i \alpha
     + \alpha \partial_i \log \bar{g}^{1/2} \; , \\
\pounds_\beta \alpha &=& \beta^i \bar{\nabla}_i \alpha
     + \alpha \bar{\nabla}_i \beta^i \; .
\end{eqnarray}

The Riemann tensor is fixed by
\begin{equation}
\left( \nabla_\alpha \nabla_\beta - \nabla_\beta \nabla_\alpha \right)
     V^\gamma = V^\delta R^\gamma\mathstrut_{\delta \alpha \beta}
\end{equation}
while the Ricci tensor is $R_{\delta \beta} \equiv
R^\gamma\mathstrut_{\delta \gamma \beta}$.

The $3+1$ decompositions of the Riemann and Ricci tensors are
\begin{eqnarray}
R_{i j k l}(g) &=& R_{i j k l}(\bar{g}) + 2 K_{i [k} K_{l] j} \; ,\\
R_{0 i j k}(g) &=& 2 N \bar{\nabla}_{[j} K_{k] i} \; ,\\
R_{0 i 0 j}(g) &=& N \left( \bpz K_{i j} + N K_{i k} K^k\mathstrut_j
     + \bar{\nabla}_i \partial_j N \right) \; .
\end{eqnarray}
One can then obtain for the Ricci tensor
\begin{eqnarray}
R_{i j}(g) &=& R_{i j}(\bar{g}) - N^{-1} \bpz K_{i j} + K K_{i j}
     - 2 K_{i k} K^k\mathstrut_j - N^{-1} \bar{\nabla}_i \partial_j N
     \; ,\\
R_{0 j}(g) &=& N \left( \partial_j K - \bar{\nabla}_h K^h\mathstrut_j
     \right) \; ,\\
R_{0 0}(g) &=& N \left( \partial_0 K - N K_{i j} K^{i j} +
     \triangle_{\bar{g}} N \right) \; ,
\end{eqnarray}
where $K\equiv K^i\mathstrut_i$ and 
$\triangle_{\bar{g}} \equiv g^{i j} \bar{\nabla}_i \bar{\nabla}_j$.
Finally, we note
\begin{equation}
G^0\mathstrut_0 = \frac{1}{2} \left( K_{i j} K^{i j} - K^2 -
     R(\bar{g}) \right) \; .
\end{equation}

\section{Every Time Slicing Is ``Harmonic''}
\label{Sec:EverySlicing}
The standard statement of the harmonic time-slicing
\index{harmonic time-slicing}
condition is, that on a $t=\mbox{const.}$ time slice,
$\bar{\partial}_0 [(-g)^{1/2} g^{00}]=0$.  (This 
is equivalent, in a coordinate basis, 
to $\partial_\mu [(-g)^{1/2} g^{\mu t}]=0$.) Friedrich observed
in~\cite{Fri85} that the right hand side of these equations could be a 
given function of $(t , x^i)$\footnote{Just as in
electrodynamics, $\nabla^\mu A_\mu = 0 \rightarrow 
\nabla^\mu {A_\mu}^{\prime} = \ell(t ,x) \neq 0$ is
perfectly acceptable as a ``Lorentz gauge'' if $\ell$ 
is known.}. (See also~\cite{CBY95}.)  Therefore, the 
standard harmonic condition expressed in 
$3+1$ form, $\bar{\partial}_0 N 
+ N^2K = 0$, can be written as
a generalized ``harmonic'' condition
\index{generalized harmonic condition}
\begin{equation}
\bar{\partial}_0 N + N^2 K = Nf \; ,
\label{GeneralHarmonic}
\end{equation}
where $f(t,x)$ is a known function.  
Specifically, introduce $\alpha(x,t)$ such that 
$\bpz \log \alpha = f$, then~(\ref{GeneralHarmonic})
becomes
\begin{equation}
\bpz N + N^2 K = N \bpz \log \alpha \; ,
\label{SpecialHarmonic}
\end{equation}
from which the identity
\begin{equation}
\bpz \log \bar{g}^{1/2} = - N K \; ,
\end{equation}
allows us to see that
\begin{equation}
N = \bar{g}^{1/2} \alpha \; .
\label{slicingfunction}
\end{equation}
We shall call $\alpha(x,t)$ the ``slicing function;''
it is a freely given scalar density of weight $-1$.

It is clear that any $N>0$ on a given time slice $t=t_0$
can be written in the form $N_{t_0}= \bar{g}^{1/2}_{t_0}
\alpha (t_0, x)$ for some $\alpha > 0$ provided that 
${g}_{ij} (t_0,x)$ is a proper riemannian metric.
\index{metric!riemannian}  Introducing 
``harmonic'' time-slicing is thus a simple matter.  It is not,
however, known at present how to construct a specific long-time
foliation from general rules telling how to specify $\alpha (t,x)$.
However, many foliations can be constructed in a ``step-by step''
fashion (numerical time steps) provided certain obvious
conditions are met. For example, an elliptic condition on $N$
can determine $\alpha(t)$ on a sequence of time slices
if the condition does not couple to variables that disturb the
characteristic directions of the hyperbolic equations.
(The same is true for the shift vector $\beta^i$.)  
Alternatively, we can try educated guesses for $\alpha(t,x)$.

As it stands,~(\ref{SpecialHarmonic}) is clearly a speed zero
(with respect to $\partial_0$) hyperbolic equation.  However,
this equation and Einstein's equations\index{Einstein equations} 
lead to a 
second order equation in space and time that propagates $N$
along the light cone.  This result brings
into sharp relief the congruence of~(\ref{SpecialHarmonic}) with
the propagation on the light cone of other variables.

The trace of $R_{i j}(g)$ gives an equation for $\bpz K$~\cite{CBY95}
\begin{equation}
\bpz K = -\triangle_{\bar{g}} N + \left[ R(\bar{g}) + K^2 - 
     R^k\mathstrut_k(g) \right] N \; ,
\label{bpzK}
\end{equation}
where $\triangle_{\bar{g}}$ in (\ref{bpzK}) denotes the Laplacian
$g^{ij} \bar{\nabla}_i \bar{\nabla}_j$.  Taking the time
derivative of~(\ref{SpecialHarmonic}) and eliminating
$\bpz K$ with~(\ref{bpzK})
shows that $N$ obeys the non-linear wave equation
\begin{equation}
\barbox_g N + R^k\mathstrut_k(g) N - R(\bar{g})N - N \bpz \log \alpha 
	+ (\bpz^2 \log \alpha)
N^{-1}=0 \;,
\label{NLWaveEQ}
\end{equation}
where we wrote our wave operator or ``d'Alembertian'' as $\barbox_g
= -(N^{-1}\bpz)^2 + \triangle_{\bar{g}}$.  The characteristic cone
of $\barbox_{g}$ is clearly the physical light cone $(c=1)$.
The equation~(\ref{NLWaveEQ}) {\em per se}
will not be used explicitly in the sequel.

Substitutions of the form $N_\lambda = \bar{g}^{\lambda/2}
\alpha_{\lambda} \; 
(\lambda > 0)$ have also been considered~\cite{FrR96}.  
However, after working out 
the wave equation analogous to~(\ref{NLWaveEQ})
that $N_\lambda$ obeys, one finds
that the local proper propagation speed of 
$N_\lambda$ is $\sqrt{\lambda}$.  This behavior may or may not
spoil the propagation of system variables other than
$N$, but if $\lambda \neq 1$ and the system is hyperbolic, one will
always find that the characteristic directions of the system will
not all be physical ones.  That is, in vacuum gravity, there will 
be some variables that propagate neither on the light cone (speed = 1)
nor along the axis parallel to $\bpz$ (speed $= 0$) which is
orthogonal to $t =$ const.  The variables not propagating in physical
directions are gauge variables and one will not have physical
criteria for their boundary values on characteristic surfaces.
On the other hand, with $\lambda = 1$, one has fulfilled 
a necessary condition that physical and gauge variables propagate 
together in the same directions.

In the following sections, whenever we consider hyperbolic systems,
we will focus on first-order symmetric (or symmetrizable) 
hyperbolic (``FOSH'')\index{FOSH}
\index{first-order symmetric hyperbolic}
equations possessing {\em only} physical
characteristic directions\index{physical characteristic directions}. 
We understand ``FOSH''\index{FOSH}
\index{first-order symmetric hyperbolic} in
{\em this restricted physical sense only} in this paper, and likewise for
other uses of the term ``hyperbolic.''

\section{Canonical Action and Equations of Motion}

Choice of the slicing function \index{slicing function}
$\alpha$ in~(\ref{slicingfunction})
is arbitrary ($\alpha > 0$). That $\alpha$ is freely chosen while $N$
must satisfy an equation of motion~(\ref{SpecialHarmonic}) 
suggests that it should be regarded as the
undetermined\footnote{The multipliers associated with gauge
freedom or, as here, with spacetime coordinate freedom, can be
freely chosen because they are not determined by physical conditions.
Hence, they are not true ``Lagrange multipliers.''}
multiplier in the canonical action principle 
\index{action principle!canonical} of
Arnowitt, Deser, and Misner (``ADM'')~\cite{ADMDir}.
Here we follow~\cite{AY98}.

To draw some lessons for the canonical formalism, let
us first express the $3+1$ evolution equations in 
their standard geometrical form (see: with zero shift~\cite{Lic44},
arbitrary lapse\index{lapse} and shift~\cite{CB56},
spacetime perspective~\cite{Yor79}):
\begin{equation}
\dot{g}_{i j} \equiv - 2 N K_{i j} \; ,
\label{gdot}
\end{equation}
\begin{equation}
\dot{K}_{i j} \equiv N \left( -R_{i j}(g) + {R}_{i j}(\bar{g})
	+ N K_{i j} - K_{i k} K^k_j 
	- N^{-1} \bar{\nabla}_i \partial_j N \right) \; ,
\label{Kdot}
\end{equation}
where $\dot{(\;)} \equiv \bpz(\; )$.

A brief look at (\ref{Kdot}) shows that forming the combination 
${\cal R}_{i j} = R_{i j}(g) - g_{i j} R^k\mathstrut_k(g)$
leads to an equation of motion for the
ADM canonical momentum
\begin{equation}
\pi^{i j} = \bar{g}^{1/2} \left( K g^{i j} - K^{i j} \right)
\end{equation}
that contains no constraints.\index{constraint} 
(In this section we choose units
in which $16\pi G = c=1$.)  Indeed, using
(\ref{gdot}) and (\ref{Kdot}), we obtain the
identity
\begin{eqnarray}
\dot{\pi}^{i j} &\equiv& N \bar{g}^{1/2} 
	\left( R(\bar{g}) g^{i j} - R^{i j}(\bar{g}) \right)
	- N \bar{g}^{-1/2} \left( 2 \pi^{i k} \pi^j\mathstrut_k 
	- \pi \pi^{i j} \right) \nonumber\\
	& & + \bar{g}^{1/2} \left( \bar{\nabla}^i \bar{\nabla}^j N
	- g^{i j} \bar{\nabla}_k \bar{\nabla}^k N \right)
	+ N \bar{g}^{1/2} \left[ {\cal R}^{i j} \right] \; .
\label{pidot}
\end{eqnarray}
{}From the identity $\dot{g}_{i j} = -2 N K_{i j}$, we have
\begin{equation}
\dot{g}_{i j} \equiv N \bar{g}^{-1/2} 
	\left( 2 \pi_{i j} - \pi g_{i j} \right) \; .
\label{gdot2}
\end{equation}

We now come to a  crucial observation.
Were the canonical equation for $\dot{\pi}^{i j}$ to be 
dictated by vanishing of the spatial part of the
Einstein tensor, $G^{i j}(g) = 0$, as it is in the conventional ADM 
analysis~\cite{ADMDir}, then the identity
\begin{equation}
G_{i j}(g) + g_{i j} G^0\mathstrut_0 (g) \equiv R_{i j}(g)
	- g_{i j} R^k\mathstrut_k(g) \equiv {\cal R}_{ij}
\label{calR}
\end{equation}
shows that a Hamiltonian constraint\index{constraint!Hamiltonian} 
term
$\sim\bar{g}^{1/2} G^0\mathstrut_0$
remains in the $\dot{\pi}^{i j}$ equation~(\ref{pidot}).
This would mean that the validity of the $\dot{\pi}^{i j}$
equation would be restricted to the subspace on which the 
Hamiltonian constraint
\index{constraint!Hamiltonian} is satisfied (i.e., vanishes).

Though the ADM derivation of the $\dot{\pi}^{i j}$
equation, found by varying $g_{i j}$ in their canonical
action\index{action} ($\beta^i$ is the shift vector)
\begin{equation}
S \left[ g, \pi ; N, \beta \right) 
	= \int d^4x \left( \pi^{i j} \dot{g}_{i j}
	- N {\cal H} \right) \; ,
\end{equation}
with $N(t,x) \; , \; \beta^i(t,x)$ and $\pi^{ij}$ held fixed,
is of course perfectly correct, another point
of view is possible.  [We are ignoring boundary terms,
a subject not of interest here, and we note that the
momentum constraint term \index{constraint!momentum}
$-\beta^i {\cal H}_i$ 
(${\cal H}_i = g^{1/2} {\cal C}_i$, ${\cal C}_i = 2 N R^0\mathstrut_i$) 
is contained in
$\pi^{i j} \dot{g}_{i j}$ 
($\dot{\left( \; \right)} \equiv \bpz $)
upon integration by parts.]
(The slicing density
\index{slicing function} $\alpha$ has also been used prominently
in the action\index{action} by Teitelboim~\cite{Tei82},
who simply set $\alpha = 1$ ($N=\bar{g}^{1/2}$), and by 
Ashtekar~\cite{Ash88,Ash87} for other purposes.)

We have explained that $\alpha$ can be regarded as a free 
undetermined multiplier while $N$ is a
{\em dynamical variable} (a conclusion also reached by
Ashtekar for other reasons~\cite{Ash88,Ash87}) 
that determines
the proper time $N \delta t$ between slices $t=t^\prime$
and $t=t^\prime+\delta t$. $N$ is determined from 
$\alpha(t,x)$ and $\bar{g}^{1/2}$ found by solving the
initial value constraint equations\index{constraint!initial value}.
(See the treatment of the constraints in the final section
of this article and in~\cite{Yor79,Yor73,OMY74a,CBY80}.)
Motivated by this viewpoint,
we alter the undetermined multiplier $N$ in the ADM
action principle\index{action principle!ADM} 
to $\alpha$, where the Hamiltonian
density $\tilde{\cal H}$ is 
(with ${\cal H} \equiv 2 \bar{g}^{1/2} G^0\mathstrut_0(g)$
being the ADM Hamiltonian density of weight $+1$)
\begin{equation}
\tilde{\cal H} \equiv \bar{g}^{1/2} {\cal H} = 
	\pi_{i j} \pi^{i j} - \frac{1}{2} \pi^2 - \bar{g} 
	R(\bar{g}) \; ,
\end{equation}
a scalar density of weight $+2$ and 
a rational function of the metric.  The action\index{action} 
becomes
\begin{equation}
S\left[\bar{g}, \pi; \alpha, \beta \right) =
	\int d^4 x \left( \pi^{i j} \dot{g}_{i j}
	- \alpha \tilde{\cal H} \right) \; .
\label{action}
\end{equation}
The modified action {\em principle}
\index{action principle!modified} for the canonical
equations that we propose in~(\ref{action}) is to vary $\pi^{i j}$ and
$g_{i j}$, with $\alpha(t,x)$ and $\beta^i(t,x)$ as
fixed undetermined multipliers. From
\begin{eqnarray}
\delta \tilde{\cal H} &=& \left( 2 \pi_{i j} - g_{i j} \pi \right)
	\delta \pi^{i j} 
	+ \left( 2 \pi^{i k} \pi^j\mathstrut_k - \pi \pi^{i j} 
	+ \bar{g} R^{i j}(\bar{g}) - \bar{g} g^{i j} R(\bar{g}) \right) 
	\delta g_{i j} 	\nonumber\\
	& & - \bar{g} \left( \bar{\nabla}^i \bar{\nabla}^j \delta g_{i j}
	- g^{i j} \bar{\nabla}_k \bar{\nabla}^k \delta g_{i j} \right) \; ,
\end{eqnarray}
we obtain the canonical equations
\begin{eqnarray}
\dot{g}_{i j} = \alpha \frac{\delta \tilde{\cal H}}{\delta \pi^{i j}}
	&=& \alpha \left( 2 \pi_{i j} - \pi g_{i j}\right)
	\equiv -2 N K_{i j} \; , \\
\label{Canonicalgdot}
\dot{\pi}^{i j} = -\alpha \frac{\delta \tilde{\cal H}}{\delta g_{i j}}
	&=& - \alpha \bar{g} \left( R^{i j}(\bar{g}) - R(\bar{g}) 
	g^{i j} \right )
	- \alpha  \left( 2 \pi^{i k} \pi^j\mathstrut_k - \pi \pi^{i j} \right)
	\nonumber\\
	& & + \bar{g} \left( \bar{\nabla}^i \bar{\nabla}^j \alpha - 
	\bar{g}^{i j} \bar{\nabla}_k \bar{\nabla}^k \alpha \right) \; .
\label{Canonicalpidot}
\end{eqnarray}
Equation (\ref{Canonicalpidot}) for $\dot{\pi}^{i j}$
is the identity (\ref{pidot}) with ${\cal R}^{i j}=0$,
which is equivalent to $R_{i j}(g)=0$. Thus, (\ref{Canonicalpidot}) is a 
``strong'' equation unlike its ADM counterpart, which requires
in addition the imposition of a constraint\index{constraint}: 
${\cal H}=0$.

In the present formulation, the canonical equations of 
motion hold everywhere on phase space with any 
parameter time $t$, a necessary condition for the issue
of ``constraint evolution''
\index{constraint evolution} even to be {\em discussed} 
in the Hamiltonian
\index{Hamiltonian dynamics} framework. (See below in 
Sect.~\ref{ContractedBianchi}.)

If we define the ``smeared'' Hamiltonian as the integral
of the Hamiltonian density,
\begin{equation}
\tilde{\cal H}_\alpha = \int d^3 x^\prime \alpha(t,x^\prime) 
	\tilde{\cal H} \;,
\end{equation}
the equation of motion for a general functional
$F[\bar{g},\pi;t,x)$ anywhere on the phase space
is 
\begin{equation}
\dot{F}\left[ \bar{g}, \pi; t, x \right) = 
	- \left\{\tilde{\cal H}_\alpha, F\right\} 
	+ \tilde{\partial}_0 F \; ,
\label{FunctionalEOM}
\end{equation}
where $\dot{\left( \; \right)}$ denotes our total time
derivative and $\tilde{\partial}_0$ is a ``partial'' 
derivative of the form $\partial_t - \pounds_\beta$
acting only on explicit spacetime dependence.  The 
Poisson bracket is
\begin{equation}
\left\{F,G\right\} = \int d^3 x 
	\left( \frac{\delta F}{\delta g_{i j}(t,x)}
	\frac{\delta G}{\delta \pi^{i j}(t,x)}
	- 
	\frac{\delta G}{\delta g_{i j}(t,x)}
	\frac{\delta F}{\delta \pi^{i j}(t,x)} \right) \; ,
\label{PoissonBracket}
\end{equation}
and one sees that 
time evolution is generated by the Hamiltonian vector field
\begin{eqnarray}
{\cal X}_{\tilde{\cal H}_\alpha} &=& \int d^3 x \left\{ 
	\alpha ( 2 \pi_{i j} - \pi g_{i j} )
	\frac{\delta}{\delta g_{i j}} 
	- [ \alpha \bar{g} ( R^{i j}(\bar{g}) 
	- R(\bar{g}) g^{i j}) \right.
\nonumber\\
	& & + \left. \alpha ( 2 \pi^{i k} \pi^j\mathstrut_k 
	- \pi \pi^{i j} )
	- \bar{g} ( \bar{\nabla}^i \bar{\nabla}^j \alpha 
	- g^{i j} \bar{\nabla}_k \bar{\nabla}^k \alpha ) ] 
	\frac{\delta}{\delta \pi^{i j}} \right\} \; .
\label{HamVecField}
\end{eqnarray}
Because it does not contain any explicit constraint\index{constraint}
dependence,~(\ref{HamVecField}) is a valid time evolution
operator on the entire phase space.
It is clear that the $\left( \dot{\bar{g}}, \dot{\pi} \right)$
equations come from~(\ref{HamVecField}) applied to the
canonical variables. The harmonic time slicing
\index{harmonic time-slicing} 
equation~(\ref{SpecialHarmonic}) results from application 
of~(\ref{HamVecField})
to $N$, and the wave equation for $N$ comes from a repeated 
application of~(\ref{HamVecField}) to~(\ref{SpecialHarmonic}).

Evolution equations for the ``constraints'' 
\index{constraint!evolution equations} are computed to be
\begin{eqnarray}
\bpz\tilde{\cal H} &=& - \left\{ \tilde{\cal H}_\alpha, 
      \tilde{\cal H} \right\}
	= \alpha \bar{g} g^{i j} \partial_i {\cal H}_j 
	+ 2 \bar{g} g^{i j} {\cal H}_i \bar{\nabla}_j \alpha \; , 
\label{newConstraint1}\\
\bpz{\cal H}_j &=& - \left\{ \tilde{\cal H}_\alpha, {\cal H}_j \right\}
	= \alpha \partial_j \tilde{\cal H} 
	+ 2 \tilde{\cal H} \partial_j \alpha \; ,
\label{newConstraint2}
\end{eqnarray}
where $\bar{\nabla}_j \alpha = \partial_j \alpha 
	+ \alpha \bar{g}^{-1/2} \partial_i \bar{g}^{1/2}$.
These are well-posed evolution 
equations for the constraints,
\index{constraint!evolution equations} and they are equivalent
to the twice-contracted Bianchi
identities\index{Bianchi identities!twice-contracted}
when ${\cal R}_{i j}=0$ or $R_{i j}=0$
(see below).

These results shed new light
on the Dirac ``algebra''\index{Dirac algebra}
of constraints~\cite{Tei73}.
It is well known that the
Dirac algebra is not the spacetime diffeomorphism
algebra.  This can be seen from the fact that
while the action~(\ref{action})\index{action} is invariant under
transformations generated by ${\cal H}_j$ and 
$\tilde{\cal H}$,~\cite{Tei77}
the equations of motion that follow from this action\index{action} are
$R_{i j}(g)=0$ even when ${\cal H}_j$ and $\tilde{\cal H}$
do not vanish.  These equations of motion are preserved by
spatial diffeomorphisms and time translations along
their flow in phase space, whereas a general spacetime
diffeomorphism applied to $R_{i j}(g)=0$ would mix in the 
constraints.\index{constraint}

A second important view of the Dirac algebra
results from the direct and beautiful dynamical
meaning of its once-smeared form.  
Equations~(\ref{newConstraint1}) and~(\ref{newConstraint2})
express {\em consistency} of the constraints\index{constraint} as a 
{\em well posed} initial-value problem. If the constraint
functions vanish in some region on an intial time slice, 
they continue to do
so under evolution by the Hamiltonian vector field into the 
domain of dependence 
of that initial region.  This mechanism
follows from the dual role of $\tilde{\cal H}$ as a 
constraint\index{constraint} and as part of the generator of
time translations of functionals of the canonical
variables anywhere on the phase space.

Let us take note that the Hamiltonian constraint
\index{constraint!Hamiltonian}
{\em per se} does not express the dynamics of 
the theory; the equation of dynamics 
is~(\ref{FunctionalEOM}).  In its ``altered'' 
role, the Hamiltonian constraint function simply
vanishes as an initial value condition
\index{constraint!initial value}, from which 
$\bar{g}^{1/2}$ is determined as in the initial value
problem.~\cite{Yor79} Then $N$ can be constructed
{}from $\alpha$. The Hamiltonian constraint,
\index{constraint!Hamiltonian} once
solved, remains so according to the results
embodied in~(\ref{newConstraint1}) and~(\ref{newConstraint2}).

\section{Contracted Bianchi Identities}
\label{ContractedBianchi}

The results on canonical dynamics that follow on using $\alpha$ 
as an undetermined multiplier are also reflected in the
manner in which the twice-contracted Bianchi identities,
\index{Bianchi identities!twice-contracted}
\begin{equation}
\nabla_\beta G^\beta\mathstrut_\alpha \equiv 0 \; ,
\label{BiID}
\end{equation}
can be written as a first-order symmetrizable hyperbolic
system~\cite{AY98}.  
(In the absence of hyperbolic form,~(\ref{BiID}) is
practically useless in providing physical equations of 
motion for the constraints\index{constraint!evolution equations} 
when they are {\em not}
satisfied.)  Likewise, this system extends to
matter (see below). (Frittelli obtained well-posedness
for~(\ref{BiID}) by other methods~\cite{Fri97}.)

We recall that the equations of motion of the canonical
momenta in vacuum are
\begin{equation}
{\cal R}_{ij} \equiv R_{ij}(g) - g_{ij} R^k\mathstrut_k(g) = 0 \; ,
\label{CalRij}
\end{equation}
while the weight zero Hamiltonian constraint
\index{constraint!Hamiltonian} is
\begin{equation}
C = 2G^0\mathstrut_0(g) = K_{ij}K^{ij} - K^2 - R(\bar{g}) = 0 \; ,
\label{C}
\end{equation}
and the weight zero one-form momentum constraint
\index{constraint!momentum} is  
\begin{equation}
C_i = 2NR^0\mathstrut_i (g) = 2\bar{\nabla}^j (K_{ij} - Kg_{ij})
= 0 \; .
\label{Ci}
\end{equation}
Recall the identity~(\ref{calR}):
\begin{equation}
G_{ij}(g) + g_{ij}G^0\mathstrut_0(g) \equiv R_{ij}(g) - g_{ij}
R^k\mathstrut_k(g) \equiv {\cal R}_{ij} \; .
\label{Gij}
\end{equation}
Combining~(\ref{CalRij}),~(\ref{C}),~(\ref{Ci}), and~(\ref{calR}) 
with~(\ref{BiID}) gives 
the twice-contracted Bianchi identities
\index{Bianchi identities!twice-contracted} as a FOSH\index{FOSH}
\index{first-order symmetric hyperbolic} system
\begin{equation}
\dot{C} - N \bar{\nabla}^j C_j \equiv 2\left( C_j \bar{\nabla}^j 
N + NKC - NK^{ij}[{\cal R}_{ij}] \right) \; ,
\label{ConWave1}
\end{equation}
\begin{equation}
\dot{C}_j - N \bar{\nabla}_j C \equiv 2 \left( C \bar{\nabla}_j N
+ \frac{1}{2} NKC_j - \bar{\nabla}^i(N[{\cal R}_{ij}]) \right) \; .
\label{ConWave2}
\end{equation} 
Substituting ${\cal H}_i = \bar{g}^{1/2} C_i$,
$\tilde{\cal H} = \bar{g} C$, and setting the equations of
motion ${\cal R}_{ij}=0$ in~(\ref{ConWave1}) and~(\ref{ConWave2}) 
yields the 
evolution equations of the unsmeared constraints
\index{constraint!evolution equations} as 
in~(\ref{newConstraint1}) and~(\ref{newConstraint2}).

Similar considerations show how to put the ``matter conservation''
equations $\nabla_\beta T^{\alpha \beta} = 0$ into well-posed form.
This was also carried out by one of us (YCB) and 
Noutchegueme~\cite{CBN86}
but the results obtained here are more immediately physical.
Unlike~\cite{CBN86}, we use the
energy density $\varepsilon = -T^0\mathstrut_0$ rather than $\rho^{00}
	(\rho^\alpha\mathstrut_\beta \equiv T^\alpha\mathstrut_\beta 
	- \frac{1}{2} 
	\delta^\alpha\mathstrut_\beta T^\mu\mathstrut_\mu)$ to obtain 
this result.  It
is clear that such a result is possible because
\begin{equation}
H_{\alpha \beta} \equiv {\kappa}^{-1} G_{\alpha \beta}(g)-T_{\alpha \beta}
\end{equation}
vanishes as Einstein's equation and in any case satisfies
$\nabla_\beta H^\beta\mathstrut_\alpha = 0$.  
We can treat $H_{\alpha \beta}$
as we did $G_{\alpha \beta}$ above.  $({\kappa} = 8\pi G \; ; \; c=1.)$
The result is nevertheless of interest as it presents the
continuity and relativistic Euler equations\index{Euler equations}
of matter in a well-posed form.

Straightforwardly expanding $\nabla_\beta T^\beta\mathstrut_0 = 0$ and 
$\nabla_\beta T^\beta\mathstrut_i = 0$  gives the continuity and Euler
equations \index{Euler equations}
(cf.~\cite{Yor79}, p. 89), with $\varepsilon \equiv -T^0\mathstrut_0$
and the matter current one-form $j_i \equiv NT^0\mathstrut_i$.  The
continuity equation is
\begin{equation}
\bpz \varepsilon + N \bar{\nabla}^i j_i = 
N(K_{ij}T^{ij} + K \varepsilon - 2j_i a^i) \; ,
\label{hatpar1}
\end{equation} 
where $a_i \equiv \bar{\nabla}_i \log N$ is the acceleration
of observers at rest in a given time-slice.  Likewise, we
find for Euler's equation\index{Euler equations}
\begin{equation}
\bpz j_i + N \bar{\nabla}_j T^j\mathstrut_i 
	= N (K^j\mathstrut_i - 
	T^j\mathstrut_i a_j - \varepsilon a_i) \; .
\label{hatpar2}
\end{equation}
The divergence term on the left side of~(\ref{hatpar2}) spoils
the well-posed FOSH\index{FOSH}
\index{first-order symmetric hyperbolic} 
form we seek.  However, if we use the
identity $G_{ij}(g) + g_{ij} G^0\mathstrut_0 (g) \equiv R_{ij}
(g) - g_{ij} R^k\mathstrut_k(g)$ and the Einstein equations
\index{Einstein equations}
${\kappa}^{-1} G_{\alpha \beta} - T_{\alpha \beta} = 0$, or
${\kappa}^{-1} R_{\alpha \beta} = \rho_{\alpha \beta}$,
we obtain $(\varepsilon = -T^0\mathstrut_0 \; , \; j_i = N T^0\mathstrut_i)$
\begin{equation}
T^j\mathstrut_i - \delta^j\mathstrut_i \varepsilon 
	= (\rho^j\mathstrut_i - \delta^j\mathstrut_i
	\rho^k\mathstrut_k) \equiv {\cal S}^j\mathstrut_i \; .
\end{equation}
Then~(\ref{hatpar1}) and~(\ref{hatpar2}) obtain well-posed
form (if ${\cal S}^j\mathstrut_i$ is assumed known),
\begin{equation}
\bpz \varepsilon + N \bar{\nabla}^i j_i = 
-2j_i \bar{\nabla}^i N + 2NK\varepsilon + NK^{ij}
{\cal S}_{ij} \; ,
\label{hatpar3}
\end{equation}
\begin{equation}
\bpz j_i + N \bar{\nabla}_i \varepsilon = 
-2\varepsilon \bar{\nabla}_i N + NKj_i - \bar{\nabla}^j
(N {\cal S}_{ij}) \; .
\label{hatpar4}
\end{equation}

By combining~(\ref{hatpar3}) plus~(\ref{Ci}), and~(\ref{hatpar4})
plus~(\ref{Gij}), we obtain expressions of gravity constraint
evolution\index{constraint!evolution equations} in the presence of matter,
\begin{equation}
\bpz C^T - N \bar{\nabla}^i C^T_i = 
2 \left( C^T_j \bar{\nabla}^j N + NKC^T - NK^{ij} ({\kappa}^{-1}
{\cal R}_{ij} - {\cal S}_{ij}) \right) \; ,
\end{equation}
\begin{equation}
\bpz C^T_j - N \bar{\nabla}_j C^T =  2 \left[ C^T
\bar{\nabla}_j N + \frac{1}{2} NKC^T_j - \bar{\nabla}^i
\left( N({\kappa}^{-1} {\cal R}_{ij} - {\cal S}_{ij}) \right)
\right]
\end{equation}
where $C^T \equiv C + 2 \varepsilon$ and $C^T_j = C_j - 2j_j$.
This is just the form we would anticipate on the basis of
Hamiltonian dynamics\index{Hamiltonian dynamics} 
and the form~(\ref{Ci}) and~(\ref{Gij})
of the vacuum constraints\index{constraint!vacuum}.  
Thus, for gravity plus a matter field,
we obtain results analogous to~(\ref{newConstraint1})
and~(\ref{newConstraint2}) for the total system. If there are no 
violations of
constraints\index{constraint violation}, 
then $C^T = 0 \; , \; C^T_j = 0$, while the
dynamical gravity equation is ${\kappa}^{-1} {\cal R}_{ij}
- {\cal S}_{ij} = 0$.

\section{Wave Equation for $K_{i j}$}

Einstein's equations,\index{Einstein equations} 
viewed mathematically as a system
of second-order partial differential equations for the metric,
do not form a hyperbolic system without modification and
are not manifestly well-posed, though, of course, physical
information does propagate at the speed of light. A 
well-posed hyperbolic system admits unique solutions 
depending continuously on the initial data and seems to be 
required for robust, stable numerical integration and
for full treatment by the methods of modern analysis, for 
example, exploitation of energy estimates. The
well-known traditional approach achieves hyperbolicity
through special coordinate choices.\footnote{The classic
second-order fully harmonic form was given in~\cite{FB52,CB56} and
discussed, for example, in~\cite{CBY80}. It will not be discussed
in this article. A FOSH form based on these equations was given
first by Fischer and Marsden in~\cite{FiM72}.}
The formulation described here permits coordinate gauge
freedom. Because these exact nonlinear theories incorporate
the constraints, they are natural starting points for 
developing gauge-invariant perturbation theory.

Consider a globally hyperbolic manifold 
$V = \Sigma \times {\mathbb R}$ with the metric as given
in the introduction. To achieve hyperbolicity for the
$3+1$ equations, we proceed as follows.

By taking a time derivative of $R_{i j}(g)$ and subtracting
appropriate spatial covariant derivatives of the momentum
constraints\index{constraint!momentum}, one of us (YCB) and 
T.~Ruggeri~\cite{CBR83} (see also~\cite{CBY95}, 
where the shift is not set to zero) obtained
an equation with a wave operator acting on the extrinsic
curvature.\index{extrinsic curvature} In vacuum, one finds
\begin{equation}
	\bpz R_{i j}(g) - \bar{\nabla}_i R_{0 j} - \bar{\nabla}_j R_{0 i}
	= N \barbox_g K_{i j} + J_{i j} + S_{i j} = 0 \; ,
\label{bpzRij}
\end{equation}
where $\barbox_g = - \left( N^{-1} \bpz \right)^2
	+ \bar{\nabla}_k \bar{\nabla}^k$, $J_{i j}$ consists
of terms at most first order in derivatives of $K_{i j}$, 
second order in derivatives of $g_{i j}$, and second order
in derivatives of $N$, and
\begin{equation}
	S_{i j} = -N^{-1} \bar{\nabla}_i \bar{\nabla}_j (\bpz N 
	+ N^2 K)\; .
\end{equation}
The term $S_{i j}$ is second order in derivatives of $K_{i j}$
and would spoil hyperbolicity of the wave operator $\barbox$
acting on $K_{i j}$. Hyperbolicity is achieved by setting
$N = \bar{g}^{1/2} \alpha (t,x)$, or
\begin{equation}
	\bpz N + N^2 K = \bar{g}^{1/2} \bpz \alpha(t,x) \;,
\end{equation}
as discussed in Sect.~\ref{Sec:EverySlicing}.
The resulting equation combined with~(\ref{gdot}) 
forms a quasi-diagonal hyperbolic
system for the metric $g_{i j}$ 
with principal operator $\bpz \barbox$. This system
can also be put in first order symmetric hyperbolic 
form~\cite{CBY95,AACbY95}, by the introduction of sufficient auxiliary
variables and by use of the equation for $R_{0 0}$ (thus
incorporating the Hamiltonian constraint
\index{constraint!Hamiltonian}). The Cauchy
data\index{Cauchy data} for the system 
(in vacuum)~\cite{CBR83,CBY95}
are
(1) $(\bar{g}, K)$ such that the constraints
	$R_{0 i} = 0$, 
$G^0\mathstrut_0 = 0$ hold on the initial slice;
\index{constraint!momentum}
\index{constraint!Hamiltonian}
\index{constraint!initial value}
(2) $\bpz K_{i j}$ such that $R_{i j}=0$ on the intial slice;
and
(3) $N>0$ arbitrary on the initial slice.
Note that the shift $\beta^k(x,t)$ is arbitrary. Using
the Bianchi identities\index{Bianchi identities}, 
one can prove~\cite{CBR83,CBY95} that this system
is fully equivalent to the Einstein equations.
\index{Einstein equations!equivalence}
The point is that quasi-diagonal Leray~\cite{leray}
hyperbolic systems have well posed Cauchy problems\index{Cauchy problem}
and therefore unique solutions for given initial data.
Because every solution of the Einstein equations
also satisfies the $\barbox K_{i j}$ equation
in particular and provides initial data for it,
uniqueness implies, conversely, that if the initial
data for the $\barbox K_{i j}$ equation are Einsteinian,
all solutions of Einstein's equations, and only these,
are captured. The restriction on the initial value
of $\bpz K_{i j}$ prevents the higher derivative from
introducing spurious unphysical solutions.

All variables propagate either with characteristic
speed zero or the speed of light. The only variables
which propagate at the speed of light have the dimensions
of curvature, and one sees that this is a theory of
propagating curvature.\index{propagating curvature} However, a FOSH
\index{first-order symmetric hyperbolic} system that 
propagates curvature is more transparent in the ``Einstein-Bianchi''
\index{Einstein equations!hyperbolic!Einstein-Bianchi}
form (next section).

In the above formulation, the shift and $\alpha$ are arbitrary.
This and our other systems (except the Einstein-Christoffel
\index{Einstein equations!hyperbolic!Einstein-Christoffel}
system in Sect.~\ref{EinsteinChristoffel}) are
manifestly spatially covariant and all time slicings
(using $\alpha$) are allowed. Spacetime covariance is
therefore present, but not completely manifest.

By taking another time derivative and adding an appropriate
derivative of $R_{0 0}$, one finds (in vacuum)~\cite{AY98}
\begin{equation}
	\bpz \bpz R_{i j} - \bpz \bar{\nabla}_i R_{0 j}
	+ \bpz \bar{\nabla}_j R_{0 i} + \bar{\nabla}_i
	\bar{\nabla}_j R_{0 0} = \bpz (N \barbox K_{i j})
	+ {\cal J}_{i j} = 0 \;,
\end{equation}
where ${\cal J}_{i j}$ consists of terms at most third
order in derivatives of $g_{i j}$ and second order in
derivatives of $K_{i j}$. Together with $\bpz g_{i j}$,
these form a system for $(\bar{g}, {K})$
which is hyperbolic non-strict in the sense of Leray-Ohya.~\cite{LeO67}
Here, the lapse\index{lapse} itself, as well as the shift, is arbitrary
$(N>0)$. The Cauchy data\index{Cauchy data} of the previous 
form (in vacuum)
must be supplemented by $\bpz \bpz K_{i j}$ such that
$\bpz R_{i j} = 0$ on the initial slice. This guarantees
that the system is fully equivalent to Einstein's theory
\index{Einstein equations!equivalence}
(except that its solutions are not in Sobolev spaces~\cite{AY98}).
This system does not have a first order symmetric hyperbolic
formulation, but has been used very effectively in perturbation
theory~\cite{AAL98} and in other applications~\cite{YCB97a,YCB97b}.

\section{Einstein-Bianchi Hyperbolic System}

To obtain a first order symmetric hyperbolic system, one can 
use the Riemann tensor of the spacetime metric. It satisfies the
Bianchi identities for the spacetime geometry
\begin{equation}
\nabla_\alpha R_{\beta \gamma \lambda \mu}
     + \nabla_\beta R_{\gamma \alpha \lambda \mu}
     + \nabla_\gamma R_{\alpha \beta \lambda \mu} \equiv 0 \; .
\end{equation}
These identities imply by contraction and use of the symmetries of the
Riemann tensor
\begin{equation}
\nabla_\alpha R^\alpha\mathstrut_{\mu \beta \gamma} +
     \nabla_\gamma R_{\beta \mu} + \nabla_\beta R_{\gamma \mu}
     \equiv 0 \; .
\end{equation}
If the Ricci tensor $R_{\alpha \beta}$ satisfies the Einstein
equations ($\kappa = c = 1$)
\begin{equation}
R_{\alpha \beta} = \rho_{\alpha \beta} \; ,
\end{equation}
then the previous identities imply the equations
\begin{equation}
\nabla_\alpha R^\alpha\mathstrut_{\mu \beta \gamma} =
     \nabla_\beta \rho_{\gamma \mu} -
     \nabla_\gamma \rho_{\beta \mu} \; .
\end{equation}

The first equations with $(\alpha \beta \gamma) = (i j k)$
and the last one with $\mu=0$ do not contain derivatives
of the Riemann tensor transverse to $M_t$. They are
considered as ``constraints'' and will be identically satisfied
(initially) in our method. They remain satisfied in an exact
integration. All detail and rigor concerning this elegant
system is given in~\cite{CBYA98,ACBY97},
to which the reader is referred.
It has 66 equations, just as do the Einstein-Ricci first
order curvature equations.

The system we are now developing~\cite{CBY97} is similar to an analogous
system obtained by H.~Friedrich~\cite{Fri96} that is based on the
Weyl tensor. The Weyl tensor system is causal but with
additional unphysical characteristics. 

We wish first to show that the remaining equations are, for $n=3$
in the vacuum case, when $g$ is given, a symmetric first order
hyperbolic system for the double two-form 
$R_{\alpha \beta \lambda \mu}$. For this purpose, following
Bel~\cite{Bel58,Bel61} we introduce two pairs of ``electric''
and ``magnetic'' space tensors associated with a spacetime
double two-form $A$,
\begin{eqnarray}
N^2 E_{i j}(g) &\equiv& A_{0 i 0 j} \\
D_{i j}(g) &\equiv& \frac{1}{4} \epsilon_{i h k} \epsilon_{j l m}
     A^{h k l m} \\
N H_{i j}(g) &\equiv& \frac{1}{2} \epsilon_{i h k} 
     A^{h k}\mathstrut_{0 j} \\
N B_{j i}(g) &\equiv& \frac{1}{2} A_{0 j}\mathstrut^{h k} \epsilon_{i h k}
\end{eqnarray}
where $\epsilon_{i j k}$ is the volume form of $\bar{g}$.
It results from the symmetry of the Riemann tensor $R$ with
respect to its first and second pairs of indices ($R$ is a
``symmetric double two-form'') that if $A \equiv R$, then $E$
and $D$ are symmetric while $H_{i j} = B_{j i}$. A useful
identity for a symmetric double two-form like $R$, with a tilde
representing the spacetime double dual, is (``Lanczos identity'')
\begin{equation}
\tilde{R}_{\alpha \beta \lambda \mu} + R_{\alpha \beta \lambda \mu}
     = C_{\alpha \lambda} \, g_{\beta \mu} -
       C_{\alpha \mu} \, g_{\beta \lambda} +
       C_{\beta \mu} \, g_{\alpha \lambda} -
       C_{\beta \lambda} \, g_{\alpha \mu} \; ,
\end{equation}
where $C_{\alpha \beta} = R_{\alpha \beta} - (1/4) g_{\alpha \beta}
R$.
It follows that when $R_{\alpha \beta} = \lambda \, g_{\alpha \beta}$,
then $E = -D$ and $H = B$. In order to avoid introducing unphysical
characteristics, and to be able to extend the treatment to the
non-vacuum case, {\em we do not use these properties in the evolution
equations}, but write them as a first order system for an arbitrary
double two-form $A$, as follows:
\begin{eqnarray}
& \nabla_0 A_{h k 0 j} + \nabla_k A_{0 h 0 j}
      - \nabla_h A_{0 k 0 j} = 0 \; , & \\
& \nabla_0 A^0\mathstrut_{i 0 j} + \nabla_h A^h_{i 0 j}
     = \nabla_0 \, \rho_{j i} - \nabla_j \, \rho_{0 i} \; ,&
\end{eqnarray}
and analogous equations with the pair $(0\, j)$ replaced by $(l\,m)$.
One obtains a first order system for the unknowns $E$, $H$, $D$,
and $B$ by using the relations inverse to the definitions
above. The principal parts of these equations, all with one
definite index fixed on $E$, $H$, $D$, and $B$, are identical
to the corresponding Maxwell equations. The characteristic matrix
of this ``Maxwell'' part of the system has determinant
$-N^6 (\xi_0 \xi^0) \, (\xi_\alpha \xi^\alpha)^2$.
The system obtained has a principal matrix consisting of 6
identical 6 by 6 blocks around the diagonal, which are symmetrizable
and hyperbolic. Hence, the system is symmetric hyperbolic, when
$g$ is a given metric such that $\bar{g}$ is properly
riemannian and $N>0$.

To relate the Riemann tensor to the metric $\bar{g}$ we use the
definition 
\begin{equation}
\bpz g_{i j} = -2 N K_{i j}
\end{equation}
and we use the $3+1$ identities given in the Introduction.
{\em Note that in this section all $\Gamma$'s are spatial.}

We next choose $N=\bar{g}^{1/2} \alpha(t,x)$. We generalize
somewhat the ideas used
by Friedrich (see~\cite{Fri96}) for the Weyl tensor to write a
symmetric hyperbolic system for $K$ and $\Gamma$, namely
we obtain equations relating $\Gamma$ and $K$, for a given
double two-form $A$, and by considering the definition of $K$ and the 
$3+1$ decomposition of the Riemann tensor,
replacing in these identities the Riemann
tensor by $A$. To deduce from this
system a symmetric hyperbolic first order system, with the
algebraic form of the harmonic gauge $N=\bar{g}^{1/2} \alpha$,
one uses the fact that in this gauge one has
\begin{equation}
\Gamma^h\mathstrut_{i h} = \partial_i \log N -
     \partial_i \log \alpha \; .
\end{equation}
We obtain
\begin{eqnarray}
\bpz \Gamma^h\mathstrut_{i j} + N \bar{\nabla}^h K_{i j}
      &=& N K_{i j} g^{h k} \left( \Gamma^m\mathstrut_{m k}
      + \partial_k \log \alpha \right) \nonumber\\
& & - 2 N K^h\mathstrut_{(i} \left( \Gamma^m\mathstrut_{j) m}
      + \partial_{j)} \log\alpha \right) \\
& & - N \left( \epsilon^k\mathstrut_{( j}\mathstrut^h B_{i) k}
      + H_{k (i} \epsilon^k\mathstrut_{j)}\mathstrut^h \right) \; ,
\nonumber
\end{eqnarray}
and
\begin{eqnarray}
\bpz K_{i j} + N \partial_h \Gamma^k \mathstrut_{i j}
     &=& N \left[ \Gamma^m\mathstrut_{i h} 
     \Gamma^h\mathstrut_{j m} - \left(
     \Gamma^h\mathstrut_{i h} + \partial_i \log \alpha \right)
     \left( \Gamma^k\mathstrut_{j k} + \partial_j \log\alpha
     \right) \right] \nonumber\\
& & - N \left( \partial_i \partial_j \log\alpha -
     \Gamma^k\mathstrut_{i j} \partial_k \log \alpha \right) \\
& & - N \left[ D_{(i j)} + E_{(i j)} - D^k\mathstrut_k g_{i j}
    - K K_{i j} \right] \; . \nonumber
\end{eqnarray}
The system obtained for $K$ and $\Gamma$ has a 
characteristic matrix composed of 6 blocks around the diagonal,
each block a 4 by 4 matrix that is symmetrizable hyperbolic,
with characteristic polynomial $-N^4 (\xi_0 \xi^0) \, 
     (\xi_\alpha \xi^\alpha)$.

The whole system for $A$, $K$, $\Gamma$, $\bar{g}$ is
symmetrizable hyperbolic, with characteristics the light cone and the
normal to $M_t$. It is somewhat involved to prove that a solution of
the constructed system satisfies the Einstein equations if the initial
data satisfy the constraints, but we can argue as follows. We consider
the vacuum case with initial data satisfying the Einstein
constraints. These initial data determine the initial values of
$\Gamma$, and also, if $\beta$ and $N$ are known at $t=0$, the
initial values of $A_{i j h m}$, $A_{j h i 0}$, $A_{i 0 j h}$ by
using the decomposition formulas. (We set A equal to the Riemann
tensor on the initial surface.) We use the Lanczos formula to
determine $A_{i0j0}$ initially. We know that our symmetrizable
hyperbolic system has one and only one solution. Because a solution of
Einstein's equations with $N=\bar{g}^{1/2} \alpha$, proved to exist in
the section on $\barbox K_{i j}$, satisfies together with its Riemann
tensor the present system and takes the same initial values, that
solution coincides with the solution of the present system in their
common domain of existence.

\section{Einstein-Christoffel System}
\label{EinsteinChristoffel}

The first-order form of the wave equation for $K_{i j}$, the
Einstein-Ricci\index{Einstein equations!hyperbolic!Einstein-Ricci} 
system~\cite{CBY95,AACbY95}, to which
we have alluded, has 66 equations, the correct number for a
curvature system as does the Einstein-Bianchi system
of the previous section. It is symmetric hyperbolic
\index{symmetric hyperbolic} and therefore
well-posed. But it is natural to ask whether there is a simpler
first-order system of fewer variables that is perhaps closer
in form to~(\ref{gdot}) and~(\ref{Kdot}). Frittelli and 
Reula~\cite{FrR94,FrR96} proposed
such a system, but their system has unphysical characteristics
and is not written fully in terms of geometric variables. Here we deduce
a different system having only physical characteristics and
expressed in geometric variables. We understand (private
communication to JWY) that James Bardeen
has likewise obtained a similar system improving that found
in~\cite{BM92,BMSS95}. 
While our derivation~\cite{AY99} can proceed systematically by direct
construction of an energy norm and the characteristic speeds,
a more heuristic derivation is indicated here from the structure
of the wave equation for $K_{i j}$. Not that in this section
all $\Gamma$'s are spatial.

In the dynamical spacetime Ricci tensor $R_{i j}(g)$,~(\ref{Kdot}),
one has a dynamical equation for the extrinsic curvature $K_{i j}$
\index{extrinsic curvature}
in terms of spatial derivatives of the spatial Christoffel
symbols. When the lapse\index{lapse} $N$ is replaced by the 
slicing density\index{slicing function}
through $\alpha \bar{g}^{1/2}$, the differentiated Christoffel
terms become
\begin{equation}
	\partial_k \Gamma^k\mathstrut_{i j}
	- \partial_j \Gamma^k\mathstrut_{i k}
	- \partial_i \Gamma^k\mathstrut_{j k} \; .
\end{equation}
This may be read as the divergence of a linear combination of
Christoffel symbols, which puts the $R_{i j}(\bar{g})$ equation 
in a form reminiscent of the structure of one of the first-order
equations for a free wave, namely
\begin{equation}
\partial_0 u + \partial^k v_k = 0\;.
\label{udot}
\end{equation}
We would have a symmetric hyperbolic system if there were 
an analog of the other equation for a wave,
\begin{equation}
\partial_0 v_k + \partial_k u = 0 \; .
\label{vdot}
\end{equation}
Some manipulation quickly leads to the conclusion that $K_{i j}$
and the Christoffel combination above are not paired in a 
symmetric hyperbolic system like $u$ and $v_k$.

Recall however, that the free wave equation $(\partial_0)^2 u
	- \partial^k \partial_k u = 0$ is obtained by taking a 
time derivative of~(\ref{udot}) and subtracting the
divergence of~(\ref{vdot}). In obtaining the wave equation
for $K_{i j}$~(\ref{bpzRij}) we have taken a time derivative of $R_{i j}$
and subtracted a (suitably symmetrized) divergence of the
momentum constraint\index{constraint!momentum} $R_{0 i}$. 
This motivates the speculation
that in gravity the ``other'' equation should be related to the
momentum constraint\index{constraint!momentum} and its sole spatial derivative should be
$\partial_k K_{i j}$.

Following this idea about the second equation leads one to 
consider an equation of the form
\begin{equation}
g_{k i} R_{j 0} + g_{k j} R_{i 0} = -\bpz f_{k i j}
	- \partial_k (N K_{i j}) + l.o._{k i j}
\label{mom1}
\end{equation}
where $l.o._{k i j}$ are lower order terms involving no
derivatives of $f_{k i j}$ or $K_{i j}$. One must choose
$f_{k i j}$ from a linear combination of spatial derivatives
of the metric. Introduce
\begin{equation}
{\cal G}_{k i j} = \partial_k g_{i j}
\label{calGkij}
\end{equation}
and use the identity
\begin{equation}
\bpz(\partial_k g_{i j}) = -\partial_k (2 N K_{i j})
\end{equation}
to find that
\begin{equation}
f_{k i j} = \frac{1}{2} {\cal G}_{k i j} -
	g_{k (i} g^{r s} \left( {\cal G}_{|r s| j)}
	- {\cal G}_{j) r s} \right)
\end{equation}
produces the correct coordinate derivatives occuring in the
momentum constraints\index{constraint!momentum}. 
The lower order terms are those terms
necessary to complete~(\ref{mom1}) into an identity
and take the form
\begin{eqnarray}
	l.o._{k i j} &=& 2 N K_{k (i} g^{r s} \left( {\cal G}_{|r s| j)}
	- {\cal G}_{j) r s} \right) \nonumber\\
& &
	+ 2 g_{k (i} \Bigl[ K_{j) m} \partial^m N - K \partial_{j)}
	N \nonumber\\
& &  + N K_{j) m} g^{r s} \Gamma^m\mathstrut_{r s}({\cal G})
	+ \frac{1}{2} N \left( {\cal G}_{j) r s} - 2 {\cal G}_{| r s | j)}
	\right) K^{r s} \Bigr] \; ,
\label{lokij}
\end{eqnarray}
where the spatial Christoffel symbols are constructed from 
${\cal G}_{i j k}$,
\begin{equation}
\Gamma_{k i j}({\cal G}) \equiv (1/2) \left( {\cal G}_{j k i}
	+ {\cal G}_{i k j} - {\cal G}_{k i j} \right) \; .
\label{bGam_def}
\end{equation}
(It is clear
that only one of the three-index symbols ${\cal G}_{k i j}$,
$\Gamma_{k i j}$, and $f_{k i j}$ is needed,
say $f_{k i j}$. The necessary algebra will not be reproduced
here.)

One then easily verifies that by expressing~(\ref{bGam_def}) in
terms of derivatives of the metric (assuming a metric compatible
\index{metric compatibility}
connection), we can manipulate it to take the form of a 
divergence of $f_{k i j}$ plus lower order terms. The
dynamical Ricci equation becomes
\begin{equation}
R_{i j} = -N^{-1} \bpz K_{i j} - \partial^k f_{k i j}
	+ l.o._{i j} \; ,
\label{Rij3}
\end{equation}
where
\begin{eqnarray}
l.o._{i j} &=& K K_{i j} - 2 K_{i k} K^k\mathstrut_j - \alpha^{-1}
	\left( \partial_i \partial_j - \Gamma^k\mathstrut_{i j}
	({\cal G}) \partial_k \right) \alpha \nonumber\\
	& & - \left( \Gamma^k\mathstrut_{k i} ({\cal G})
	+ \alpha^{-1} \partial_i \alpha \right)
	\left( \Gamma^m\mathstrut_{m j}({\cal G})
	+ \alpha^{-1} \partial_j \alpha \right) \\
	& & + 2 \Gamma^k\mathstrut_{m k}({\cal G})
	\Gamma^m\mathstrut_{i j}({\cal G})
	- \Gamma^k\mathstrut_{m j}({\cal G})
	\Gamma^m_{i k}({\cal G}) \nonumber\\
	& & + g^{k r} g^{s m} \left[ {\cal G}_{k r s} f_{m i j}
	+ {\cal G}_{k m (i} {\cal G}_{j) r s}
	- {\cal G}_{k r s} {\cal G}_{(i j) m} \right]\; .\nonumber
\end{eqnarray}
Together with~(\ref{gdot}),~(\ref{mom1}) 
and~(\ref{Rij3}) constitute a symmetric hyperbolic system for the
evolution of $g_{i j}$, $K_{i j}$ and $f_{k i j}$. 

Note
that once $f_{k i j}$ or, equivalently, ${\cal G}_{k i j}$
are introduced as variables, the relation~(\ref{bGam_def}) becomes
an initial condition and does not {\em a priori} hold
for all time. Equation~(\ref{mom1}) can be
related to metric compatibility \index{metric compatibility}
by putting it in the form
\begin{eqnarray}
4 g_{k (i} R_{j) 0} &=& \bpz {\cal G}_{k i j} + \partial_k
	\left( 2 N K_{i j} \right) \\
& & - 4 g_{k (i} N \bar{\nabla}^m \left( K_{j) m} - g_{j) m} K \right)
	\; . \nonumber
\label{R0k_gg}
\end{eqnarray}
Here, one sees that if the momentum constraint
\index{constraint!momentum} is satisfied
for all time, then
\begin{equation}
{\cal G}_{k i j} = 2 \Gamma_{(i j) k} =
	\partial_k g_{i j}
\end{equation}
and the connection is metric compatible.\index{metric compatibility} 
If the momentum
constraint\index{constraint!momentum} 
is violated, metric compatibility
\index{metric compatibility} is sacrificed.
This shows the price paid to achieve a symmetric
hyperbolic system that is close to the canonical equations:
the momentum constraints\index{constraint!momentum} 
become dynamical, and metric
compatibility is lost if the latter are violated. 
\index{metric compatibility}

\section{Conformal ``Thin Sandwich'' Data for the Initial Value
Problem}
\label{ConformalTS}

The standard approach to the initial value problem
\index{initial value problem} is the ``conformal
method,''\index{initial value problem!conformal method} 
the fundamental rudiments of which were introduced by
Lichnerowicz~\cite{Lic44}.  The essentially complete form was developed by
two of us (YCB and JWY), see~\cite{Yor73,CBY80}. 
Basic theorems were obtained
by us and by O'Murchadha~\cite{OMY73,OMY74a,OMY74b}, 
and Isenberg and Moncrief~\cite{IsM94}.
The older approach concentrates (in the vacuum
case to which we restrict ourselves) on the construction of the
spatial metric\index{metric!spatial} 
$g_{ij} = \psi^4 \gamma_{ij}$ and the traceless
part $A^{ij}= \psi^{-10} \lambda^{ij}$
of the extrinsic curvature\index{extrinsic curvature}, where $K_{ij} = A_{ij}
+ \frac{1}{3} g_{ij}K$.  Here, $\gamma_{ij}$ is a proper
riemannian metric\index{metric!riemannian} 
given freely, $\lambda^{ij}$ is constructed
by a tensor decomposition method~\cite{Yor73,Yor74}, 
and $K$ is given freely
(not conformally transformed).  Note that one may as well assume 
$\det(\gamma_{ij}) = 1$ because only the conformal equivalence 
class of the metric\index{metric!conformal equivalence class} 
matters:  the entire method is ``conformally
covariant.''\index{conformal covariance}\\

N.B.:  In Sect.~\ref{ConformalTS}, only spatial metrics
\index{metric!spatial} will be used.  Therefore,
{\em all overbars are dropped in this final section}.\\

Here, we discuss a new interpretation of the four Einstein vacuum
\index{Einstein equations!vacuum}
initial-value constraints\index{constraint!initial value}.  
(The presence of matter would add
nothing new to the analysis.)  Partly in the spirit of a ``thin
sandwich'' viewpoint, this approach is based on prescribing the
{\em conformal} metric\index{metric!conformal}~[1] 
on each of two nearby spacelike 
hypersurfaces (``time slices'' $t=t^{\prime}$  and $t = t^{\prime}
+ \delta t$) that make a ``thin sandwich'' (TS).  {\em Essential use
is made of the understanding of the slicing function in general
relativity}.  The new formulation could prove useful both 
conceptually and in practice, as a way to construct initial data
in which one has a hold on the input data different from that in the
currently accepted approach.  The new approach allows us to 
{\em derive} from its dynamical and metrical foundations the 
important scaling law ${A}^{ij} = \psi^{-10} \lambda^{ij}$ for the
traceless part of the extrinsic curvature.
\index{extrinsic curvature}  This rule is simply
postulated in the one-hypersurface approach.

The constraint equations\index{constraint!vacuum} 
on $\Sigma$ are, in vacuum,
\begin{equation}
\nabla_j (K^{ij} - K g^{ij}) = 0 \; ,
\end{equation}
\begin{equation}
R(g) - K_{ij}K^{ij} + K^2 = 0
\end{equation}
where $R(g)$ is the spatial scalar curvature of $g_{ij},\; \nabla_j$
is the Levi-Civita connection of $g_{ij}$; and $K$ is the trace of
$K_{ij}$, also called the ``mean curvature'' of the slice.  

The time derivative of the spatial metric
\index{metric!spatial} $g_{ij}$ is related to
$K_{ij}, \; N$, and the shift vector $\beta^i$ by
\begin{equation}
\partial_t g_{ij} \equiv -2NK_{ij} + (\nabla_i \beta_j + \nabla_j
\beta_i) \; ,
\label{gdotKNbeta}
\end{equation}
where $\beta_j = g_{j i} \beta^i$. The fixed spatial coordinates
$x$ of a point on the ``second'' hypersurface, as evaluated
on the ``first'' hypersurface, are displaced by 
$\beta^i(x) \delta t$ with respect to those on
the first hypersurface, with an orthogonal link from the first 
to the second surface as a fiducial reference:
$\beta_i = \frac{\partial}{\partial t}
* \frac{\partial}{\partial x^i}$,
where $*$ is the physical spacetime inner product of the
indicated natural basis four-vectors. The essentially arbitrary
direction of $\frac{\partial}{\partial t}$
is why $N(x)$ and $\beta^i(x)$ appear in the TS formulation.
In contrast, the tensor $K_{i j}$ is always determined
by the behavior of the unit normal on one slice and therefore
does not possess the kinematical freedom, i.e., the gauge variance, 
of $\frac{\partial}{\partial t}$. Therefore, $N$
and $\beta^i$ do not appear in the one-hypersurface IVP for
$(\Sigma, {g}, {K})$.

Turning now to the conformal metrics\index{metric!conformal} 
in the IVP, we recall that
two metrics $g_{i j}$ and $\gamma_{i j}$ are conformally 
equivalent\index{conformal equivalence} 
if and only if there is a scalar $\psi > 0$ such
that $g_{i j} = \psi^4 \gamma_{i j}$. The conformally
invariant representative of the entire conformal equivalence
class\index{metric!conformal equivalence class}, 
in three dimensions, is the weight $(-2/3)$ 
unit-determinant ``conformal metric''\index{metric!conformal} 
$\hat{g}_{i j} = g^{-1/3} g_{i j} = \gamma^{-1/3} \gamma_{i j}$
with $g = \mbox{det}(g_{i j})$ and 
$\gamma = \mbox{det}(\gamma_{i j})$. Note particularly
that for any small perturbation, $g^{i j} \delta \hat{g}{i j} = 0$.
We will use the important relation
\begin{equation}
g^{i j} \partial_t \hat{g}_{i j} 
     = \gamma^{i j} \partial_t \hat{g}_{i j}
     = \hat{g}^{i j} \partial_t \hat{g}_{i j}
     = 0 \; .
\label{ImportantRelation}
\end{equation}

In the following, rather than use the mathematical apparatus 
associated with conformally weighted objects such as 
$\hat{g}_{i j}$, we find it simpler to use ordinary scalars
and tensors to the same effect. Thus, let the role of 
$\hat{g}_{i j}$ on the first surface be played by a given metric
$\gamma_{i j}$ such that the physical metric that satisfies
the constraints is $g_{i j} = \psi^4 \gamma_{i j}$ for
some scalar $\psi > 0$. (This corresponds to ``dressing''
the initial unimodular conformal metric\index{metric!conformal} 
$\hat{g}_{i j}$ with 
the correct determinant factor $g^{1/3} = \psi^4 \gamma^{1/3}$.
This process does not alter the conformal equivalence class of
the metric.)\index{metric!conformal equivalence class} 
The role of the conformal metric\index{metric!conformal} on the second
surface is played by the metric 
$\gamma^\prime_{i j} = \gamma_{i j} + u_{i j} \delta t$,
where, in keeping with (\ref{ImportantRelation}), the 
velocity tensor $u_{i j} = \partial_t \gamma_{i j}$
is chosen such that
\begin{equation}
\gamma^{i j} u_{i j} 
     = \gamma^{i j} \partial_t \gamma_{i j} = 0 \; .
\end{equation}
Then, to first order in $\delta t$, $\gamma^\prime_{i j}$ and 
$\gamma_{i j}$ have equal determinants, as desired;
but $\gamma_{i j}$ and $\gamma^\prime_{i j}$ are not
in the same conformal equivalence class in general.
\index{metric!conformal equivalence class}

We now examine the relation between the covariant derivative
operators $D_i$ of $\gamma_{i j}$ and $\nabla_i$ of
$g_{i j}$. The relation is determined by
\begin{equation}
\Gamma^i\mathstrut_{j k} (g) = \Gamma^i\mathstrut_{j k}(\gamma)
     + 2 \psi^{-1} \left( 2 \delta^i\mathstrut_{( j} 
     \partial_{k )} \psi - \gamma^{i l} \gamma_{j k}
     \partial_l \psi \right) \; ,
\label{CDrelation}
\end{equation}
{}from which follows the scalar curvature relation first
used in an initial-value problem by Lichnerowicz~\cite{Lic},
\begin{equation}
R(g) = \psi^{-4} R(\gamma) - 8 \psi^{-5} 
     \triangle_\gamma\psi \;,
\end{equation}
where $\triangle_\gamma \equiv \gamma^{k l} D_k D_l \psi$
is the scalar Laplacian associated with 
$\gamma_{i j}$.

Next, we solve~(\ref{gdotKNbeta}) for its traceless part
\begin{equation}
\partial_t g_{i j} - \frac{1}{3} g_{i j} g^{k l}
	\partial_t g_{k l} \equiv V_{i j}
	= -2 N A_{i j} + (L_g \beta)_{i j}
\label{TracelessPart}
\end{equation}
with $A_{i j} \equiv K_{i j} - (1/3) K g_{i j}$
and
\begin{equation}
(L_g \beta)_{i j} \equiv \nabla_i \beta_j + \nabla_j \beta_i
	- (2/3) g_{i j} \nabla^k \beta_k \;.
\label{Lgbeta}
\end{equation}

Expression (\ref{Lgbeta}) vanishes, for non-vanishing $\beta^i$,
if and only if $g_{i j}$ admits a conformal Killing vector
$\beta^i = k^i$. Clearly, $k^i$ would also be a conformal Killing 
vector of $\gamma_{i j}$, or of any metric conformally
equivalent\index{conformal equivalence} 
to $g_{i j}$, with no scaling of $k^i$. This teaches
us that $\beta^i$ does not scale. That $\beta^i$ does not scale
also follows because, as generator of a spatial diffeomorphism,
it is not a dynamical variable. We take the latter ``rule'' 
as a matter of principle.

It is clear in (\ref{TracelessPart}) that the left hand side
$u_{i j}$ satisfies $u_{i j} = \psi^4 V_{i j}$ because the
terms in $\dot\psi$ cancel out. Furthermore, a straightforward
calculation shows that
\begin{equation}
(L_g \beta)_{i j} = \psi^4 \left[ L_\gamma (\psi^{-4} \beta)\right]_{i j}
\, ; \quad
(L_g \beta)^{i j} = \psi^4 (L_\gamma \beta)^{i j}\; ,
\end{equation}
where $\psi^{-4} \beta_j = \gamma_{i j} \beta^i$.
Next, we note that the lapse function $N$\index{lapse}
has essential non-trivial
conformal behavior. This is a new element in the IVP analysis.
The slicing function 
$\alpha(t,x)>0$ can replace the lapse function $N$,\index{lapse}
\begin{equation}
N = g^{1/2} \alpha \; .
\end{equation}
(The treatment here extends the one in~\cite{Yor99} in 
a simple but interesting way.)
We have concluded that $\alpha$ is not a dynamical variable
and therefore does not scale. Furthermore, without loss of generality,
we can set $\mbox{det} (\gamma_{i j}) = 1$ and thus
$g^{1/2} = \psi^6$. Then
\begin{equation}
N = \psi^6 \alpha \; .
\end{equation}
Finally, we fix $K$ and require that it does not scale, 
as in the standard treatment of
the IVP~\cite{Yor72}. 
This step is absolutely essential for geometric
consistency, as we shall see.

Next we solve (\ref{CDrelation}) for $A^{i j}$, using the scaling
rules established above, and find
\begin{equation}
A^{i j} = \psi^{-10} \left\{ \frac{1}{2\alpha} \left[
	(L_\gamma \beta)^{i j} - u^{i j}\right] \right\}\;.
\end{equation}
The momentum constraint\index{constraint!momentum} becomes
\begin{equation}
D_j \left[ \frac{1}{2\alpha} (L_\gamma \beta)^{i j}\right]
	= D_j \left[ \frac{1}{2 \alpha} u^{i j}\right]
	+ \frac{2}{3} \psi^6 \gamma^{i j} \partial_j K \; ,
\label{MomConD}
\end{equation}
while the Hamiltonian constraint becomes~\cite{Yor73}
\index{constraint!Hamiltonian}
\begin{equation}
8 \triangle_\gamma \psi - R(\gamma) \psi + (\gamma_{i k} \gamma_{j l})
	A^{i j} A^{k l} \psi^{-7} - (2/3) K^2 \psi^5 = 0\;.
\label{HamConA}
\end{equation}
The unknowns $(\psi,\beta)$ obey equations of the same form as 
do the conformal scalar potential $\phi$ and the vector
potential $W^i$ in
the standard analysis~\cite{Yor73,CBY80}, but {\em no tensor splittings} are
required. Further, (\ref{MomConD}) and (\ref{HamConA}) are
coupled in only one direction when $K = \mbox{const}$.

Now we note two interesting consequences of this approach. First
we see that from $N = \psi^6 \alpha$, we have identically
\begin{equation}
N = g^{1/2} \alpha
\end{equation}
as a consequence of the method. Therefore, time slices $t$ 
and $t+\delta t$ have a relation that is manifestly ``harmonic:''
\begin{equation}
\bpz N + N^2 K = N \bpz \log \alpha\;,
\end{equation}
a result that is fully consistent with our previous discussions
and requiring that $K$ be a fixed, non-scaling, variable.

Finally, we can establish the final relationships between the
full riemannian metrics\index{metric!riemannian} $g_{i j}(t)$ and 
$g^\prime_{i j} = g_{i j}(t+\delta t)$ on the two manifestly
harmonically related slices $t$ and $t+\delta t$. As 
in~(\ref{gdotKNbeta}),
\begin{equation}
\partial_t g_{i j} = \partial_t (\psi^4 \gamma_{i j})
	= g_{i k} g_{j l} \left[ -2 N (A^{k l} + \frac{1}{3}
	g^{k l} K) + (\nabla^k \beta^l + \nabla^l \beta^k)\right] \;.
\label{ggprimeRelation}
\end{equation}
Working out~(\ref{ggprimeRelation}) gives
\begin{eqnarray}
\partial_t g_{i j} &=& \psi^4 \left[ u_{i j} + \gamma_{i j}
	\partial_t (\psi \log \psi) \right] \nonumber\\
	&=& V_{i j} + g_{i j} \partial_t (\psi \log \psi) \; ,
\end{eqnarray}
where
\begin{eqnarray}
\partial_t (\psi \log \psi) &=& \frac{2}{3} \left( D_k \beta^k
	+ 6 \beta^k \partial_k \log \psi - \alpha K \psi^6 \right)
\nonumber\\
	&=& \partial_t (g/\gamma)^{1/2} = \partial_t (g)^{1/3}
	= \frac{2}{3} \left( \nabla_k \beta^k - N K \right) \; .
\end{eqnarray}
Hence, $\partial_t \psi$ and $\partial_t g_{i j}$ are fully
determined and we note that the no-scaling rules for $\beta^k$ 
and $K$ were essential.

\section*{Acknowledgments}
This paper is based on a lecture given by James W. York Jr.
at the $2^\mathrm{nd}$ Samos meeting, 1 September, 1998.

The authors thank the supporters of the $2^{\mathrm{nd}}$ Samos
Meeting and its hosts, especially
Spiros Cotsakis, for their warm hospitality and the beautiful
setting of the meeting. They thank
Mark Hannam and Sarah and Mark Rupright for help in preparing
the manuscript. AA and JWY acknowledge support from National
Science Foundation grant PHY-9413207.






\begin{thebibliography}{99}

\bibitem{Wheeler} 
{Wheeler~J.A. (1964) Geometrodynamics and the issue of the final state. 
In: DeWitt~C., DeWitt~B. (Eds.) Relativity, Groups, and Topology. 
Gordon and Breach, New York, 317-520.}
\bibitem{ADMDir}
{Arnowitt~R., Deser~S., Misner~C.W. (1962) The dynamics of general 
relativity. In: Witten~L. (Ed.) Gravitation. Wiley, New York, 227-265.}
\bibitem{Dir58}
{Dirac~P.A.M. (1958) The theory of gravitation in Hamiltonian form. 
Proc Roy Soc A246:333-343.}
\bibitem{Dir59} 
{Dirac~P.A.M. (1959) Fixation of coordinates in the Hamiltonian 
theory of gravitation. Phys Rev 114:924-930.}
\bibitem{Lic} 
{Lichnerowicz~A. (1939) Probl\'emes Globaux en Mecanique
Relativiste. Hermann, Paris.}
\bibitem{FB52} 
{Choquet (Four\`es)-Bruhat~Y. (1952) Th\'eor\`em d'existence pour 
certains syst\`emes d'equations aux d\'eriv\'ees partielles non 
lin\'eaires. Acta Math 88:141-225.}
\bibitem{CB56} 
{Choquet (Four\'es)-Bruhat~Y. (1956) Sur l'Integration des 
\'equations de la relativit\'e g\'en\'erale.
J Rat Mechanics and Anal 5:951-966.}
\bibitem{Lic44} 
{Lichnerowicz~A. (1944) L'int\'egration des \'equations de la 
gravitation relativiste et le probl\`eme des $n$ corps. J Math 
Pures et Appl 23:37-63.}
\bibitem{CBY80} 
{Choquet-Bruhat~Y., York~J.W. (1980) The Cauchy Problem. In: Held~A. 
(Ed.) General Relativity and Gravitation, I. Plenum, New York, 99-172.}
\bibitem{AY99}
{Anderson~A. and York~J.W. (1999) Fixing Einstein's equations.
Phys Rev Lett 81:4384-4387.}
\bibitem{AACbY95} 
{Abrahams~A., Anderson~A., Choquet-Bruhat~Y., York~J.W. (1995) 
Einstein and Yang-Mills theories in hyperbolic form
without gauge fixing. Phys Rev Lett 75:3377-3381.}
\bibitem{AACbY96} 
{Abrahams~A., Anderson~A., Choquet-Bruhat~Y., York~J.W. (1996) 
A non-strictly hyperbolic system for the Einstein equations
with arbitrary lapse and shift.
C.R. Acad Sci Paris S\'erie IIb 323:835-841.}
\bibitem{AY96} 
{Abrahams~A.,York~J.W. (1997) 3+1 general relativity
in hyperbolic form. In:  Marck~J-A., 
Lasota~J-P. (Eds.) Relativistic Astrophysics and
Gravitational Radiation. North Holland, Amsterdam, 179-190.}
\bibitem{AACbY97} 
{Abrahams~A., Anderson~A., Choquet-Bruhat~Y., 
York~J.W. (1997) Geometrical hyperbolic systems for general 
relativity and gauge theories. Class Quantum Grav 14:A9-A22.}
\bibitem{AACBY97b}
{Abrahams~A., Anderson~A., Choquet-Bruhat~Y.,
York~J.W. (1998) Hyperbolic formulation of general relativity. 
In: Olinto~A.V., Frieman~J.A., Schramm~D.N. (Eds.) Proc
1996 Texas Symposium on Relativistic Astrophysics. World Scientific, 
Singapore, 601-603.}
\bibitem{BM92} 
{Bona~C., Mass\'o~J. (1992) Hyperbolic evolution system for 
numerical relativity. Phys Rev Lett 68:1097-1099.}
\bibitem{BMSS95} 
{Bona~C., Mass\'o~J., Seidel~E., Stela~J. (1995) New formalism for 
numerical relativity. Phys Rev Lett 75:600-603.}
\bibitem{CBR83} 
{Choquet-Bruhat~Y., Ruggeri~T. (1983) Hyperbolicity of the 3+1 
system of Einstein equations. Commun Math Phys 89:269-275.}
\bibitem{CBY95} 
{Choquet-Bruhat~Y., York~J.W. (1995) Geometrical well posed systems 
for the Einstein equations.
C.R. Acad Sci Paris, S\'erie I t321:1089-1095.}
\bibitem{CBY96} 
{Choquet-Bruhat~Y., York~J.W. (1996) Mixed Elliptic
and Hyperbolic Systems for the Einstein Equations. In: 
Ferrarese~G. (Ed.) Gravitation, Electromagnetism and Geometric
Structures. Pythagora Editrice, Bologna, Italy, 55-74.}
\bibitem{CBY97} 
{Choquet-Bruhat~Y., York~J.W. (1997) Well posed reduced
systems for the Einstein equations. Banach Center Publications, 
Part 1 41:119-131.}
\bibitem{CBYA98} 
{Choquet-Bruhat~Y., York~J.W., Anderson~A. (1998) Curvature-based 
hyperbolic systems for general relativity. To appear in: Piran~T. 
(Ed.) Proc 1997 Marcel Grossmann Meeting, gr-qc/9802027.}
\bibitem{FiM72} 
{Fischer~A., Marsden~J. The Einstein evolution
equations as a first-order quasi-linear symmetric hyperbolic system. I.
Commun Math Phys 28:1-38.}
\bibitem{Fri85} 
{Friedrich~H. (1985) On the hyperbolicity of Einstein and other 
gauge field-equations. Commun Math Phys 100:525-543.}
\bibitem{Fri96} 
{Friedrich~H. (1996) Hyperbolic reductions for Einstein's
equations.
Class Quantum Grav 13:1451-1459.}
\bibitem{FrR94} 
{Frittelli~S., Reula~O. (1994) On the Newtonian
limit of general relativity.
Commun Math Phys 166:221-235.}
\bibitem{FrR96} 
{Frittelli~S., Reula~O. (1996) First-order symmetric hyperbolic 
Einstein equations with arbitrary gauge.
Phys Rev Lett 76:4667-4670.}
\bibitem{VPE96} 
{van~Putten~M.H.P.M., Eardley~D.M. (1996) Nonlinear wave equations 
for relativity.
Phys Rev D53:3056-3063.}
\bibitem{AY98}
{Anderson~A. and York~J.W. (1998) Hamiltonian time evolution for
general relativity. Phys Rev Lett 81:1154-1157.}
\bibitem{Yor79} 
{York~J.W. (1979) Kinematics and dynamics of general relativity. 
In: Smarr~L. (Ed.) Sources of Gravitational Radiation, Cambridge 
Univ Press, Cambridge, 83-126.}
\bibitem{Tei82} 
{Teitelboim~C. (1982) Quantum mechanics of the gravitational field. 
Phys Rev D25:3159-3179.}
\bibitem{Ash88}
{Ashtekar~A. (1988) New Perspectives in Canonical Gravity. 
Bibliopolis, Naples.}
\bibitem{Ash87} 
{Ashtekar~A. (1987) New Hamiltonian formulation of
general relativity. 
Phys Rev D36:1587-1602.}
\bibitem{Yor73} 
{York~J.W. (1973) Conformally invariant orthogonal decomposition of 
symmetric tensors on riemannian manifolds and the
initial-value problem of general relativity. J Math Phys 14:456-464.}
\bibitem{OMY74a} 
{O'Murchadha~N., York~J.W. (1974) Initial-value problem of general 
relativity. I. General formulation and physical
interpretation. Phys Rev D10:428-436.}
\bibitem{Tei73} 
{Teitelboim~C. (1973) How commutators of constraints reflect the 
spacetime structure. Ann Phys (NY) 79:542-557.}
\bibitem{Tei77} 
{Teitelboim~C. (1977) Supergravity and square root of constraints. 
Phys Rev Lett 38:1106-1110.}
\bibitem{Fri97} 
{Friedrich~H. (1991) On the global existence and the 
asymptotic-behaviour of solutions to the Einstein-Maxwell-Yang-Mills 
equations. J Diff Geom 34:275-345.}
\bibitem{CBN86} 
{Choquet-Bruhat~Y., Noutchegueme~N. (1986) Syst\`eme hyperbolique 
pour les \'equations d'Einstein avec sources. C.R. Acad Sc
Paris, S\'erie I t303:259-263.}
\bibitem{leray} 
{Leray~J. (1952) Hyperbolic Differential Equations. I.A.S lecture 
notes, Princeton.}
\bibitem{LeO67}
{Leray~J., Ohya~Y. (1967) \'Equations et syst\`emes non-lin\'earies, 
hyperboliques nonstricts. Math Ann 170:167-205.}
\bibitem{AAL98}
{Anderson~A., Abrahams~A., Lea~C. (1998) Curvature-based
gauge-invariant 
perturbation theory for gravity: a new paradigm. Phys Rev D5806:4015.}
\bibitem{YCB97a}
{Choquet-Bruhat Y. (1997) High frequency oscillations of Einstein
geometry. In: Ibragimov N., Mahomed F. (Eds.) Modern Group Analysis.
World Scientific, Singapore, 17-34.}
\bibitem{YCB97b}
{Choquet-Bruhat Y., Greco A. (1997) Interactions of gravitational
and fluid waves. Circ Mat di Palermo 38:112-121.}
\bibitem{OMY73} 
{O'Murchada~N., York~J.W. (1973) Existence and uniqueness of solutions 
of the Hamiltonian constraint of general
relativity on compact manifolds. J Math Phys 14:1551-1557.}
\bibitem{OMY74b} 
{O'Murchadha~N., York~J.W. (1974) Initial-value problem of general 
relativity. II. Stability of solutions of the initial-value
equations. Phys Rev D10:437-446.}
\bibitem{IsM94} 
{Isenberg~J., Moncrief~V. (1994) Constraint equations with non-constant
mean curvature.  In: Flato~M., Kerner~R., Lichnerowicz~A. (Eds.) 
Physics on Manifolds. Kluwer, Dordrecht, The Netherlands, 295-301.}
\bibitem{Yor74} 
{York~J.W. (1974) Covariant decompositions of symmetric tensors 
in the theory of gravitation. Ann Inst Henri Poincar\'{e}, 
Section A 21:319-332.}
\bibitem{Yor99} 
{York~J.W. (1999) Conformal ``thin-sandwich'' data for the 
initial-value problem of general relativity. Phys Rev Lett 82:1350-1353.}
\bibitem{Yor72} 
{York~J.W. (1972) Role of conformal three-geometry in the 
dynamics of gravitation. Phys Rev Lett 28:1082-1085.}
\bibitem{ACBY97}
{Anderson~A., Choquet-Bruhat~Y., York~J.W. (1997) Einstein-Bianchi 
hyperbolic system for general relativity.
Topol Meth Nonlinear Anal  10:353-373.}
\bibitem{Bel58}
{Bel~L. (1958) D\'efinition d'une densit\'e d'\'energie et 
d'un \'etat de radiation totale. C.R. Acad Sci Paris 246:3105-3108.}
\bibitem{Bel61}
{Bel~L. (1961) L' identitie de Bianchi. Th\'ese,
University of Paris, Paris.}

\end{thebibliography}
\end{document}